\documentclass{caosp302}
\usepackage{float}
\usepackage{graphicx}

\articleNo{123}
\pubyear{2007}
\volume{35}
\volnumber{3}
\firstpage{1}
\received{October, 2007}
\accepted{August 28, 2005}

\begin{document}

%%%%%%%%%%%%%%%%%%%%%%%%%%%%%%%%%%%%%%%%%%%%%%%%%%%%%%%%%%%%%%%%%%%%%%%%%%%%%
%              R U N N I N G   P A G E   H E A D I N G S                     
% Odd page headings (except for the title page) are produced automatically
% and contain the title. If, and only if, the title of your article is too
% long the running head is ommitted in the printout; you can make your own
% running title by using the \htitle command, putting the shortened title
% between the curly brackets. \htitle should also be used when the
% subtitle is present: \htitle offers you a way how to include it into the
% headings. If you wish to see how it works simply remove the % sign from
% the beginnig of that line.
%
% Unlike the \htitle command, the \hauthor command is compulsory. It is
% used to produce even page headings and contains the names of the authors
% of an article.  All authors must be listed here, if possible. When
% authors' list is too long, you can abbreviate it by using "{\it et
% al.}". Authors' names are given in the form: initial(s) of the author's
% first name and surname. Authors are separated by a "," (comma) sign and
% the last one by "and".
%%%%%%%%%%%%%%%%%%%%%%%%%%%%%%%%%%%%%%%%%%%%%%%%%%%%%%%%%%%%%%%%%%%%%%%%%%%%%
\htitle{Spots on the surface of HgMn stars}
\hauthor{S.\,Hubrig, J. F.\,Gonz\'alez and R.\,Arlt}
%\hauthor{L.\,Neslu\v{s}an {\it et al.}}

%%%%%%%%%%%%%%%%%%%%%%%%%%%%%%%%%%%%%%%%%%%%%%%%%%%%%%%%%%%%%%%%%%%%%%%%%%%%%
%                       T I T L E                                            
% Capital letters in the title are only used at the beginning of the
% names. Don't end the title by a "." (dot)
%%%%%%%%%%%%%%%%%%%%%%%%%%%%%%%%%%%%%%%%%%%%%%%%%%%%%%%%%%%%%%%%%%%%%%%%%%%%%
\title{Spots on the surface of HgMn stars: Clues to the origin of Hg and Mn peculiarities}

%%%%%%%%%%%%%%%%%%%%%%%%%%%%%%%%%%%%%%%%%%%%%%%%%%%%%%%%%%%%%%%%%%%%%%%%%%%%%
%                       S U B T I T L E                                      
% You can use the subtitle, with the command \subtitle similar to the
% \title command.
%%%%%%%%%%%%%%%%%%%%%%%%%%%%%%%%%%%%%%%%%%%%%%%%%%%%%%%%%%%%%%%%%%%%%%%%%%%%%

%%%%%%%%%%%%%%%%%%%%%%%%%%%%%%%%%%%%%%%%%%%%%%%%%%%%%%%%%%%%%%%%%%%%%%%%%%%%%
%                   A U T H O R  N A M E S                                   
% Authors' names are separated by the \and commmand and their institutes
% are assigned by the \inst{n} command.
%
% When the name contains "Slovak" letters L,d,t,l with a caron, use an
% \additional softl, etc. command (examples given in the last table of
% this document) to produce typographically correct accented characters.
%%%%%%%%%%%%%%%%%%%%%%%%%%%%%%%%%%%%%%%%%%%%%%%%%%%%%%%%%%%%%%%%%%%%%%%%%%%%%
\author{
        S.\,Hubrig\inst{1}
      \and 
        J. F.\,Gonz\'alez\inst{2}
      \and 
        R.\,Arlt\inst{3}
       }

%%%%%%%%%%%%%%%%%%%%%%%%%%%%%%%%%%%%%%%%%%%%%%%%%%%%%%%%%%%%%%%%%%%%%%%%%%%%%
%           I N S T I T U T E S'  A D D R E S S E S                          
% The affiliation of authors is generated by the \institute command, the
% \and command being again used to separate individual addresses.
% The following commands may be used for the following three institutes:   
%               \lomnica        for      AsU SAV, Tatranska Lomnica          
%               \blava          for      AsU SAV, Bratislava                 
%               \ondrejov       for      AsU CAV, Ondrejov                   
%
% The given postal address must be complete in order to facilitate our
% editorial work. Moreover, you can add your e-mail address, using the
% \email command.
%%%%%%%%%%%%%%%%%%%%%%%%%%%%%%%%%%%%%%%%%%%%%%%%%%%%%%%%%%%%%%%%%%%%%%%%%%%%%
\institute{
           ESO, Casilla 19001, Santiago 19, Chile \email{shubrig@eso.org}
         \and 
           Complejo Astron\'omico El Leoncito, Casilla 467, 5400 San Juan, Argentina \\
           %\email{fgonzalez@casleo.gov.ar}
         \and 
           Astrophysikalisches Institut Potsdam, An der Sternwarte 16, Germany \\
           %\email{rarlt@aip.de}
          }

%%%%%%%%%%%%%%%%%%%%%%%%%%%%%%%%%%%%%%%%%%%%%%%%%%%%%%%%%%%%%%%%%%%%%%%%%%%%%
%                        D A T E / R E C E I V E D                          
% Date inserted here will be the date when your paper was received The
% format is: month (not abbreviated), day, year.
%%%%%%%%%%%%%%%%%%%%%%%%%%%%%%%%%%%%%%%%%%%%%%%%%%%%%%%%%%%%%%%%%%%%%%%%%%%%%
\date{October 8, 2007}
%\date{March 10, 2003}

%%%%%%%%%%%%%%%%%%%%%%%%%%%%%%%%%%%%%%%%%%%%%%%%%%%%%%%%%%%%%%%%%%%%%%%%%%%%%
%                        M A K E T I T L E
% The beginning part (title, author(s), etc.) of your article must be
% closed by the \maketitle command.
%%%%%%%%%%%%%%%%%%%%%%%%%%%%%%%%%%%%%%%%%%%%%%%%%%%%%%%%%%%%%%%%%%%%%%%%%%%%%
\maketitle

%%%%%%%%%%%%%%%%%%%%%%%%%%%%%%%%%%%%%%%%%%%%%%%%%%%%%%%%%%%%%%%%%%%%%%%%%%%%%
%                        A B S T R A C T,  K E Y W O R D S                   
% Here it is shown how to write an abstract.  Keywords should be placed
% within the "abstract" environment using the command \keywords and they
% should be selected from the thesaurus from Astron.  Astrophys.
% Abstracts. They must be separated from each other by -- (two dashes).
%%%%%%%%%%%%%%%%%%%%%%%%%%%%%%%%%%%%%%%%%%%%%%%%%%%%%%%%%%%%%%%%%%%%%%%%%%%%%
\begin{abstract}
%Using UVES at the VLT and FEROS at the ESO 2.2\,m telescope, we carried out a survey of 
%a large sample of HgMn 
%stars to search for spectral variability caused by an inhomogeneous 
%distribution of various elements on the surface of these stars.
The important result achieved in our recent study of 
a large sample of HgMn stars using UVES at the VLT and FEROS at the ESO 2.2\,m telescope 
is the finding that most HgMn stars exhibit
spectral variability of various chemical elements, proving
that the presence of an inhomogeneous distribution on the surface of these stars is a 
rather
common characteristics and not a rare phenomenon.
\end{abstract}

%%%%%%%%%%%%%%%%%%%%%%%%%%%%%%%%%%%%%%%%%%%%%%%%%%%%%%%%%%%%%%%%%%%%%%%%%%%%%
%                       S E C T I O N I N G                                  
% Any section starts with the command \section as shown below, with the
% title in Initial Capitals and lowercase only. Do not number the sections
% - let LaTeX do that for you - and do not end them by a "." (dot).
%
% The (sub)section titles are typeset in boldface; so, if working in the
% mathematics mode in (sub)section titles, you must use \boldmath and 
% enclose it into curly brackets, e.g. "{\bolmath $R^{2}$}".
%%%%%%%%%%%%%%%%%%%%%%%%%%%%%%%%%%%%%%%%%%%%%%%%%%%%%%%%%%%%%%%%%%%%%%%%%%%%%
%\section{Introduction}
%%%%%%%%%%%%%%%%%%%%%%%%%%%%%%%%%%%%%%%%%%%%%%%%%%%%%%%%%%%%%%%%%%%%%%%%%%%%%
%                       L A B E L                                            
% The label command is very convenient for you when referring to sections,
% subsections,..., tables, figures as well as to equations (see commands
% \ref and \pageref). In the case of figure and/or table environments the
% \label command should always be put after the \caption command to
% preserve proper numbering. When using the \label command the file must
% be compiled twice to get proper cross-references.
%%%%%%%%%%%%%%%%%%%%%%%%%%%%%%%%%%%%%%%%%%%%%%%%%%%%%%%%%%%%%%%%%%%%%%%%%%%%%
%\label{Inhomogeneous chemical abundance distribution and magnetic fields}
\section{Inhomogeneous chemical abundance distribution and magnetic fields}
%Typically, inhomogeneous chemical abundance distributions are observed only on the 
%surface of magnetic chemically peculiar stars with large-scale organized 
%magnetic fields. 
%In these stars the abundance distribution of certain elements is 
%non-uniform and non-symmetric with respect to the rotation axis.
%The presence of  magnetic fields in the atmospheres of HgMn stars
%has been studied by Mathys \& Hubrig (1995),
%Hubrig \& Castelli (2001),
%and more recently by Hubrig et al.\ (2006a) who detected a weak 
%longitudinal magnetic field in four HgMn stars. The structure of the measured field in 
%HgMn stars must be, however, sufficiently tangled so that it does not produce a strong 
%net observable circular polarization signature.
%It is intriguing that from the survey of the Ninth Catalogue of Spectroscopic Binary Orbits
%(Pourbaix et al.\ 2004, A\&A 424, 272) limited to systems brighter that V = 7, it appears 
%that only very few normal B8 and B9 stars are known to be members of SBs with periods 
%between 3 and 20~days, 
%while the majority of SB systems contain a HgMn star as a primary.
The fact that many late B-type SB systems contain a HgMn star as a primary 
%(Pourbaix et al.\ 2004)
points out 
that it is very likely that most normal late B-type stars formed in close 
binary systems are HgMn stars (Hubrig \& Mathys 1996).
From a survey of HgMn stars in close SBs, Hubrig \& Mathys (1995)
suggested that some chemical elements might be inhomogeneously
distributed on the surface, with, in particular, preferential concentration
of Mn around the rotation poles and of Hg along the equator. 
In a recent study of the star AR\,Aur which is the only known eclipsing binary with a HgMn
primary we concluded that certain elements are very likely concentrated in a fractured 
ring along the rotational equator (Hubrig et al.\ 2006a).
Also our recent survey of a large sample of HgMn 
stars using UVES at the VLT and FEROS at the ESO 2.2\,m telescope to search for spectral 
variability caused by an inhomogeneous distribution of various elements on the surface of 
these stars revealed that most HgMn stars exhibit
spectral variability of various chemical elements (Gonz\'alez et al. 2007, in preparation).
Typically, inhomogeneous chemical abundance distributions are observed only on the 
surface of magnetic chemically peculiar stars with large-scale organized 
magnetic fields.
Weak magnetic fields in the atmospheres of HgMn stars
have been detected by Mathys \& Hubrig (1995),
Hubrig \& Castelli (2001),
and more recently by Hubrig et al.\ (2006b).
The structure of the measured field in HgMn stars is expected to be, however,
sufficiently tangled so that it does not produce a strong 
net observable circular polarization signature

\begin{figure}
%\centerline{\includegraphics[width=0.45\textwidth,angle=0,clip=]{spsurf_hgmn_3x2.eps}}
\centerline{\includegraphics[width=0.48\textwidth,angle=0,clip=]{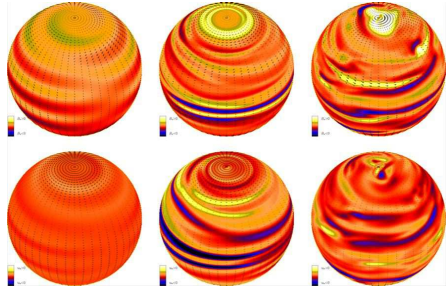}}
\caption{Radial magnetic field (upper row) and 
velocity (lower row) on the surface of a radiative star during the 
onset of the MRI
(with steps of roughly 20\,Myr).
}
%magneto-rotational instability. 
\label{fig:he3}
\end{figure}

\section{Discussion}
The role that magnetic fields possibly play in the development of anomalies in 
%HgMn stars, which mostly appear in 
binary systems, has 
never been critically tested by astrophysical dynamos. 
%This is a fundamental question 
%whose answer is essential for the understanding of the physical processes taking place 
%in HgMn stars and, more generally, during the formation and evolution of B stars.
%A scenario how a magnetic field can be built up in binary systems has been presented 
%some time ago by 
Hubrig (1998) suggested that the tidal torque varying 
with depth and latitude in a star induces differential rotation. Differential rotation 
in a radiative star can, however, be prone to magneto-rotational instability (MRI) 
(e.g., Arlt et al.\ 2003). Recent magnetohydrodynamical simulations revealed 
a distinct structure for the magnetic field topology similar to the fractured elemental rings 
observed on the surface of HgMn stars (Fig.~1). 
Complex surface patterns can be obtained from the nonlinear,
non-axisymmetric evolution of the MRI. The combination of 
differential rotation and a poloidal magnetic field was 
studied numerically by the spherical MHD code of Hollerbach 
(2000). The initial 
model differential rotation was hydrodynamically stable 
(Taylor-Proudman flow), but the introduction of a magnetic 
field excites the MRI.
%The model setup starts with a differential rotation with
%Omega decreasing from core to surface by 84\% with axis 
%distance and a homogeneous vertical magnetic field with a 
%small nonaxisymmetric perturbation. While the setup looks 
%artificial at first glance, it is actually not very important, 
%since the results are obtained after a turbulent state has 
%reorganised the system. 
The flows and fields resulting from the instability efficiently redistribute angular 
momentum and deliver a uniformly rotating star after about 10-100\,Myr.
% (Arlt et al.\2003). 
%B and A stars may thus be in the process of this redistribution. 
The presented typical 
patterns of the velocity and the magnetic field on the 
surface of the star may as well be an indication for element 
redistribution on (or in) the star.
%The results presented here
%suggest new directions for investigations to solve the question of the origin
%of abundance anomalies in B-type stars with HgMn peculiarity.

%\begin{figure}
%\includegraphics[width=5.5cm,angle=0,clip=]{sr2.ps}
%\includegraphics[width=4.0cm,angle=0,clip=]{fig5.ps}
%\caption{Left: Preliminary modelling of the abundance distribution of Sr~II on the surface
%of AR~Aur.Right: Magnetic field (left) and 
%velocity (right) on the surface of a radiative star during the 
%onset of the MRI.
%%%magneto-rotational instability. 
%The time laps between the rows is roughly 10\,Myr}
%\label{fig:he3}
%\end{figure}
%\begin{figure}
%\centerline{\includegraphics[width=5.5cm,angle=0,clip=]{sr2.ps}}
%\caption{Preliminary modelling of the abundance distribution of Sr~II on the surface
%of AR~Aur.}
%%using the direct Doppler Imaging method (Hubrig et al.\ 2006, MNRAS 371, 1953).}
%\label{fig:he3}
%\end{figure}

{}

\begin{thebibliography}{}

\article{Arlt, R., Hollerbach, R., R\"udiger, G.}{2003}{\aaa}{401}{1087}
\article{Hollerbach, R.}{2000}{Int.\ J.\ Numer.\ Meth.\ Fluids}{32}{773}
\article{Hubrig, S., Mathys, G.}{1995}{Com.\ Ap}{18}{167}
\article{Hubrig, S., Mathys, G.}{1996}{\aaas}{120}{457}
\article{Hubrig, S.}{1998}{COSKA}{27}{296}
\article{Hubrig, S., Castelli, F.}{2001}{\aaa}{375}{963}
\article{Hubrig, S., Gonz\'alez, J.F., Savanov, I.,  Sch\"oller, M., Ageorges, N., Cowley, C.R., Wolff, B.}{2006a}{\mnras}{371}{1953}
\article{Hubrig, S., North, P., Sch\"oller, M., Mathys, G.}{2006b}{\an}{327}{289}

%\bibitem{} Pourbaix,D., Tokovinin, A. A., Batten, A. H., Fekel, F. C., Hartkopf, W. I., 
%Levato, H., Morrell, N. I., Torres, G., Udry, S., 2004, \aaa \ {\bf 424}, 727

%\bibitem{}Harrison, G. R. 1939, {\it MIT Wavelength Tables},
%(New York: Wiley)

%\bibitem{}Hartoog, M. R. 1979, \apj, {\bf 231}, 161

%\bibitem{}Hartoog, M. R., and Cowley, A. P. 1979, \apj, {\it 228}, 229

%\bibitem{}Heber, U. 1991, in {\it Evolution of Stars: The
%Photospheric Abundance Connection}, ed. G. Michaud and A.
%Tutukov, IAU Symp. {\bf 145}, 363.

%\bibitem{}Hubrig, S., Castelli, F., and Mathys, G. 1999, \aaa,
%{\bf 341}, 190

%\bibitem{}Hubrig, S., Nesvacil, N., Scholler, M., and Mathys, G.,\aaa, {\bf 440}, L37

%\bibitem{}Iglesias, L., and Velasco, R. 1964, {\it Pub.
%Inst. Opt. Madrid}, No. 23.

%\bibitem{}Kurtz, D. W., Elkin, V. G., and Mathys, G. 2007,
%\aaa, {\bf 380}, 741

%\bibitem{}Leckrone, D. S., Johansson, S., Kalus, G.,
%Wahlgren, G. M., Brage, T., and Profitt, C. R. 1996,\apj, {\bf 462}, 937

%\bibitem{}McKellar, A. 1948, {\it Pub. Dom. Ap. Obs.}, {\bf 7},395

%\bibitem{}Moore, C. E. 1945, {\it Cont. Princeton Univ. Obs.},No. 20
\end{thebibliography}
\end{document}